\documentclass[aps,prb,reprint,groupedaddress]{revtex4-1}
\usepackage[pdftex]{graphicx,hyperref}
\usepackage{amsmath}
\usepackage{amssymb}
\usepackage{amsfonts}
\usepackage{dcolumn,physics}
\allowdisplaybreaks
\usepackage{bm,comment}
\usepackage{color}
\newcommand{\ch}[1]{#1}
\newcommand{\RM}[1]{#1}
\begin{document}
\preprint{}
\title{Effective spin model with anisotropic exchange interactions
for the spin-orbit coupled Hubbard model at half-filling}
\author{Ryo Makuta}
  \email{makuta-r@g.ecc.u-tokyo.ac.jp}
\author{Chisa Hotta}
   \email{chisa@phys.c.u-tokyo.ac.jp}
     \affiliation{Department of Basic Science, University of Tokyo, 3-8-1 Komaba, Meguro, Tokyo 153-8902, Japan}
\date{\today}
\begin{abstract}
Spin-orbit coupling (SOC) in noncentrosymmetric materials is the source of incommensurate magnetic structures.
In semiconductors, it drives the Rashba spin splitting and spin momentum locking,
while in magnetic insulators based on transition metals, it induces anisotropic spin exchange interactions,
like \ch{the} Dzyaloshinskii-Moriya (DM) interaction which \ch{drives} chiral magnetism and skyrmion formation.
Here, we establish a direct connection between SOC and spin exchange interactions
by deriving an effective spin model from the SOC Hubbard model at half-filling.
Using a strong-coupling expansion up to fourth order by including the full set of terms, 
we identify Heisenberg, 
Ising-like, and ring exchange interactions, as well as a variety of four-body terms for realistic Hubbard parameters, 
which impose strong constraints on the relative strengths of the spin interactions. 
Our spin model shows excellent agreement in energy with the SOC Hubbard model 
down $U/t\sim 5$ near the metal-insulator transition point, 
\ch{providing insight to which kind of magnetic interactions 
relevant across this regime are responsible for the emergence of complex magnetic textures.}
\end{abstract}
\maketitle
\section{Introduction}
In noncentrosymmetric semiconducting materials with broken inversion symmetry,
spin-orbit coupling (SOC) plays a crucial role in
imparting a topological character to the energy bands \cite{Rashba1960}.
This has brought many paths to manipulate electronic or magnetic degrees of freedom,
leading to phenomena such as Rashba spin splitting, spin-momentum locking\cite{Bychkov1984},
spin Hall effects\cite{Hirsch1999} and the Edelstein effect \cite{Edelstein1990}.
\ch{On top of these reciprocal-space physics,} SOC also drives the formation of real-space magnetic textures,
such as skyrmions, which are expected to be robust against external perturbations due to
their topological stability \cite{Nagaosa2013}.
\par
In magnetic insulators realized under strong electronic correlations,
the SOC of magnetic ions is \ch{converted to} anisotropic spin exchange interactions, enriching the quantum magnetism.
The manifestation of SOC due to broken-inversion symmetry is the
Dzyaloshinskii-Moriya (DM) interaction\cite{Dzyaloshinsky1958,Moriya1960,Kaplan1983,Shekhtman1993}.
Indeed, it is an important source of the thermal Hall effect of magnons,
while its amplitude is smaller than
the Heisenberg interactions \ch{by orders of} magnitudes in standard magnets
and it has played only a secondary role in determining the ground-state nature.
Nonetheless, there are a series of studies trying to describe the skyrmionic state by \ch{classical magnetic models}
with DM interactions\cite{Rossler2006,Muhlbauer2009,Neubauer2009,Yu2010,Yu2011,Seki2012,Tokunaga2015}.
These models mostly rely on phenomenologies, where the origin of these interactions remains unclear except that
they are allowed by the symmetry.
Furthermore, many of the skyrmions discussed there are observed as metallic ones in laboratories,
indicating a pronounced quantum fluctuation effect and the fragility of the classical picture
as they are likely to be in the parameter range closer to the weak coupling regime.
\par
The motivation of this study is thus to clarify the interrelationship between the electronic model with SOC
and its effective quantum spin model for the Mott insulating phase in the strong-to-intermediate coupling region.
In the Hubbard model with substantial SOC, the authors have found a variety of skyrmionic phases
in the intermediate interaction strength including \ch{metallic ones}, 
where the energy bands carry finite Chern numbers\cite{Makuta2024}. 
Although the simplest spin model in the strong coupling limit at the lowest order
is known to include the DM interaction term, 
a more precise evaluation of the model including higher-order terms
that applies to the weaker coupling region is needed. 
Although some studies have partially discussed the higher-order effective spin model mainly by assuming specific forms of the fourth-order exchange interaction
\cite{Paul2020,Hirschberger2021,Matteo2023,Mitra_2024}, 
no previous work has derived the complete set of such terms in the presence of SOC 
to systematically examine which of them are dominant. 
\par
Contrastingly, the anisotropic spin-exchange interactions in $4d, 5d$ and $4f$ insulators have been
systematically evaluated.
Their magnetic degrees of freedom are typically carried by the pseudo spin-1/2 of a Kramers doublet
formed by the interplay of crystal field splitting, Coulomb interactions, and SOC.
The exchange interactions reflect the spatial anisotropy of orbitals and spins and the lattice they live on,
which, even when the inversion symmetry is unbroken, generates $\Gamma$ and Kitaev terms that are anisotropic
both in real space and in spin space\cite{Rau2018,Jackeli2009}.
\par
In \S.II we explain the SOC Hubbard model, and in \S. III, perform the strong coupling expansion up to fourth order. 
Tables in \S.III provide the full set of terms of the effective spin Hamiltonian, and although technical, 
\ch{the terms serve as general forms} that are broadly applicable to models with various lattice geometries. 
We examine the nature of the effective spin model in \S. IV and discuss its implication in \S.V.

\begin{figure*}
  \includegraphics[width=17cm]{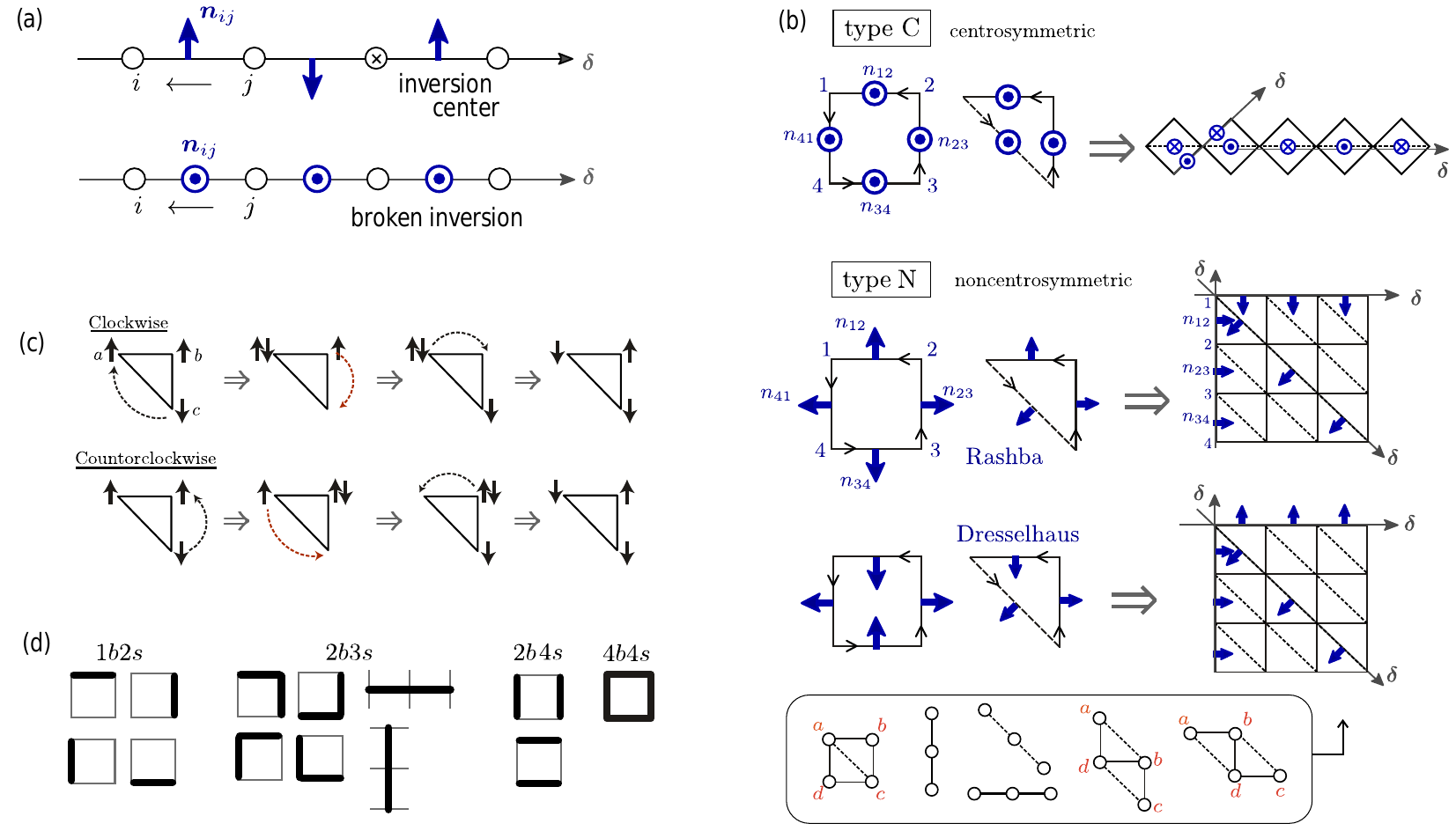}
  \caption{(a) Schematic illustration of the arrangement of SOC vector $\bm n_{ij}$
  for the systems with and without spatial inversion symmetry, 
  \RM{when the electrons travel in the $-\delta$ direction ($j\rightarrow i$). }
  (b) Unit plaquette/triangle of the fourth/third order perturbation processes  
  and the corresponding lattice structure consisting of these units. 
  Type-C and N refer to the centrosymmetric and noncentrosymmetric types of arrangements of $\bm n_{ij}$,
  \RM{for which the direction of electronic hopping is indicated by arrows on the edge bonds. }
  \ch{For anticlockwise hopping $1\rightarrow2\rightarrow3\rightarrow 4$, 
  $n_{12}=n_{23}=n_{34}=n_{41}$ for type-C, and $n_{12}\perp n_{23}$ (anticlockwise) for type-N. 
  When hopping along vertical straight bond (see those of the lattice), $-\delta$, $\bm n_{ij}$ of type-C/N follow 
  those of upper/lower configurations in panel(a). }
  For type-N there are Rashba and Dresselhaus type symmetries depending on the more detailed crystal symmetries. 
  In the actual calculation of type-N, we use the Rashba type SOC. 
  The right panel shows the square lattice with \ch{dashed} diagonal bonds 
  (when nonzero, forming a triangular lattice), obtained for each type. 
  (c) Examples of the third-order perturbation processes that appear in pairs of clockwise and
  counterclockwise types, which give the same matrix element with different signs.
  (d) Classification of fourth-order processes labeled according to the number of bonds,
  $(1b, 2b, 4b)$, and sites, $(2s, 3s, 4s)$ that participate in the process. 
  \ch{Part of the $2b3s$ process forming straight line cannot be simply accommodated inside the unit plaquette. 
   However, all these units independently apply to the processes whenever these units are found as part of the lattice. 
   On the bottom panel of (b), we show an example of how we assign 
   the three types of plaquette units $\langle abcd\rangle$, and also three-adjacent sites that contribute $2b3s$. }
  }
  \label{f1}
\end{figure*}
\section{Model and method}
\subsection{Spin-orbit coupled Hubbard model}
We consider a single-orbital Hubbard model at half-filling with SOC,
whose Hamiltonian is given as
\begin{align}
\hat{\mathcal{H}}=&
-\sum_{\langle i,j\rangle} \{\bm{c}_{i}^\dagger \big( t +i\lambda
(\bm{n}_{ij}\cdot\bm{\sigma})\big) \bm{c}_{j} + \mathrm{h.c.}\}
 +U \sum_{j}  \hat{n}_{j\uparrow}\hat{n}_{j\downarrow},
\label{eq:ham}
\end{align}
where $\langle i,j\rangle$ runs over all pairs of nearest neighboring sites,
$\bm c_{i}^\dagger= (c_{i \uparrow}^\dagger,c_{i \downarrow}^\dagger)$ is the creation operator of up and down spin electrons,
$n_{i}$ is an electron number operator, and $U$ is on-site interaction. 
The spin-dependent hopping integral $\lambda$ in Eq.(\ref{eq:ham}) originates from the atomic SOC of ions
that represent the sites\cite{Witczak2014,Nakai2022}.
Because the SOC couples the kinetic momentum of electron with the spin momentum, 
the hopping term associated with SOC is antisymmetric. 
Namely, the sign of this term is opposite 
between the electron hops in the $-\bm \delta$-direction ($j\rightarrow i$) as in Fig.~\ref{f1}(a)  
and in the $+\bm \delta$ direction, 
which is encoded in the directional dependence of the vector, 
$\bm{n}_{ij}=-\bm{n}_{ji}$\cite{Zheng2015}. 
In addition, the relative orientations of $\bm n_{ij}$'s on the lattice bonds 
are restricted by the crystal symmetry of the related material. 
When the global inversion symmetry is kept, 
$\bm n_{ij}$'s have \ch{staggered} configuration when travelling along the certain direction. 
which is realized in the centrosymmetric crystals, 
For the noncentrosymmetric cases with broken inversion symmetry, $\bm n_{ij}$ points in the same direction. 
This distinction predetermined in the Hamiltonian critically impacts the electronic and magnetic properties, 
as the spin splitting of energy bands occurs only in the \ch{latter} noncentrosymmetric materials. 
This kind of spin splitting can generate a variety of incommensurate magnetic structures
including spin-density-wave, vortex, spirals, and skyrmions\cite{Kawano2023,Makuta2024}. 
Here we combine this $\lambda$-term with the $t$-term and express them using the SU(2) gauge field;
\begin{equation}
t+i \lambda (\bm{n}_{ij}\cdot\bm{\sigma})= t_{\mathrm{eff}} \mathrm{e}^{i(\theta/2)\bm{n}_{ij}\cdot\bm{\sigma}},
\label{eq:teff}
\end{equation}
where $\bm{\sigma}=(\sigma^{x},\sigma^{y},\sigma^{z})$ is the Pauli matrix,
$t_{\mathrm{eff}}=\sqrt{t^{2}+\lambda^{2}}$ and $\theta=2\mathrm{atan2}(t,\lambda)$.
This term implies that the electrons change their spin orientations when
hopping from site $j$ to site $i$ by the angle $\theta$ about the axis $\bm{n}_{ij}$.

\subsection{Perturbation Theory}
We derive the effective spin-1/2 Hamiltonian by the perturbation from
the $U/t_{\rm eff}\rightarrow \infty$ limit of Eq.(\ref{eq:ham}) at half-filling up to fourth order. 
In the degenerate perturbation theory, 
we divide the Hilbert space into lowest and higher energy sectors, roughly separated by $U$, 
and apply the Schrieffer-Wolff transformation\cite{Schrieffer1966} 
that zero-outs the mixing between the two sectors up to fourth order. 
The number of sites that joins the $n$-th order process in terms of $t_{\rm eff}$ 
is maximally $n$, which needs to form a closed $n$-site loop,  
and thus we consider the plaquette unit shown in Fig.~\ref{f1}(b) 
as a minimal and sufficient unit to derive the effective Hamiltonian 
\footnote{Here, we also need to consider a triangular unit that may appear at $n=3$, 
however, we see shortly that these terms becomes exactly zero. 
Notice that if we consider a square lattice with only nearest neighbor, 
there is no hopping paths that form such triangular unit, so that we do not even need to 
consider such processes.  
}.
\par
The total dimension of the four electron states spanned on four sites is 
\ch{${\cal V}={}_{2n}C_{n}=70$} with $n=4$, and among them the Mott insulating states with one electron per site 
belong to the lowest energy sector which has ${\cal V}_p=2^4=16$ fold spin degeneracy. 
To be precise, we apply a so-called {\it local Schrieffer-Wolff transformation}, that converts the 
${\cal V}\times {\cal V}$ matrix $\hat H$ into 
the block diagonal form, $({\cal V}_p\times{\cal V}_p) \oplus ({\cal V}_q\times {\cal V}_q)$ 
with ${\cal V}_q={\cal V}-{\cal V}_p=54$, given as $e^{S} {\hat H} e^{-S}$, 
by the block-off-diagonal matrix $S=\sum_{n=1}^4 S_n$ with $S_n^\dagger = -S_n$. 
The form of $S_n$ is determined by demanding it as a sum of local interactions 
up to $n$ spins, and these interactions are locally block-off-diagonal on the 
subset of spins they act on. 
This transformation will account for {\it all the processes up to fourth order}, 
and for this choice of $S$, we {\it uniquely} derive the $16\times 16$ matrix representation of the effective Hamiltonian 
for the lowest energy sector\cite{Bravyi2011}.  
This effective Hamiltonian is expressed using the spin-1/2 operators $\bm S_\alpha$ with
site indices $\alpha=a,b,c,d$ (see Fig.~\ref{f1}(b)). 
\par
By the above-mentioned transformation, we derive the effective spin-1/2 Hamiltonian defined on a plaquette. 
It consists of second- and fourth-order terms. 
The second-order Hamiltonian, ${\cal H}^{(2)}$, arises from processes where electrons hop 
twice and includes only two-body exchange interactions. 
The third-order contributions cancel out (we see shortly in Fig.~\ref{f1}(c)).
The fourth-order Hamiltonian, ${\cal H}^{(4)}$, originates from processes 
involving four electron hoppings, which can be classified into four categories
(Fig.~\ref{f1}(d)); When only one bond is involved ($1b2s$), 
the contribution is to a two-body term. 
The process $(2b3s)$ also yield two-body terms, but between next-nearest-neighbor spins. 
The processes $(2b4s)$ cancel out, 
and finally, the $(4b4s)$ processes, intrinsic to the fourth order, generate genuine four-body interactions.
A complete set of these terms will be presented in Tables I and II.
\par
Once the effective Hamiltonian is derived on a plaquette unit, 
one can reconstruct spin models for a wide variety of lattice geometries. 
\ch{
The way how to make use our results of perturbation to explicitly obtain  
the effective Hamiltonian is as follows: \\
(1) Take a look at a complete set of effective spin interactions in Tables I and II, 
which we provide after running a Schrieffer-Wolff calculation to 
the SOC Hubbard model of type-C and type-N at half-filling up to fourth order. 
These processes are classified according to the geometries like $(4b4s)$ in the Tables. 
\\
(2) Specify a lattice structure for example in Fig.~\ref{f1}(b), 
a square lattice, a triangular lattice (realized by adding diagonal bonds to the square lattice, as illustrated), 
or a chain of corner-shared square unit, and define the Hamiltonian Eq.(\ref{eq:ham}) 
by properly setting the directions of $\bm n_{ij}$'s that fits to either type-C or N at your choice. 
\\
(3) Write down all the units in Fig.~\ref{f1}(d) that appear on the lattice under consideration 
without duplication. 
For example, $N\times N$ square lattice with diagonal bond and with periodic boundaries 
include three types of $\langle a b c d\rangle$ unit plaquettes, 
number $N$ for each, and three types of adjacent three-site straight bonds, number $N$ for each. 
\\
(4) Spin Hamiltonian is given as a sum of all corresponding interactions 
in the Table for all these units. 
\\
Unless one selects only particular terms among those in the Table, 
the above (1)-(4) is to be done numerically based on the data we supply.  
}
\par
Regarding the types of SOC, we consider two representative cases, centrosymmetric and noncentrosymmetric systems. 
The former is denoted as type-C, which has the \ch{staggered} $\bm{n}_{ij}$, 
where we set $\bm{n}_{ij}$ to point in the same direction perpendicular to the square plane 
when travelling along the sites $d\rightarrow c\rightarrow b\rightarrow a$. 
This type of SOC is observed for the centrosymmetric materials like Cu(1,3-bdc) 
having corner-sharing kagome structure,
which is considered to be the cause of the phase transition at low temperature\cite{Zheng2015}.
The latter is type-N, where $\bm{n}_{ij}$ points in-plane and aligned uniformly 
when travelling along the same direction. 
It corresponds to the Rashba-type SOC found in quasi-2D noncentrosymmetric materials
such as InAlAs/InGaAs\cite{Nitta1997}, LaAlO$_3$/SrTiO$_3$\cite{Caviglia2010,Nakamura2012}, Bi$_2$Se$_3$\cite{Zhang2009}, BiTeI\cite{Ishizaka2011} and heavy metals/alloys\cite{LaShell1996,Ast2007}. 
For the unit plaquette, $\bm{n}_{ij}$ rotates counterclockwise when traveling along the path
shown in Fig.~\ref{f1}(b). 
The Dresselhaus SOC also breaks the inversion symmetry, and $\bm{n}_{ij}$ rotates clockwise. 
The way how $\bm{n}_{ij}$ is arranged on a square lattice with diagonal hopping, 
or equivalently an anisotropic triangular lattice, 
is shown in the right panels for both type-C and type-N. 
The effective spin-1/2 Hamiltonian for the Dresselhaus-type is straightforwardly obtained 
from that of the Rashba-type by the local gauge transformation (local spin-axis rotation) of effective spins. 
Therefore, we only explicitly calculate the Rashba-one. 

\par
\section{Plaquette spin-1/2 Hamiltonian}

\begin{table*}[t]
\caption{Constituents of the fourth order Hamiltonian ${\cal H}^{(4)}$ for type-C with inversion symmetry. 
$c$ and $s$ included in the coefficients ${\cal J}_\eta$ 
denote $\cos(\theta/2)$ and $\sin(\theta/2)$, respectively.
The energy unit is ${t_{\rm eff}^4}{U^3}$, see Eq.(\ref{eq:4th}).}
\begin{tabular}{lllrc}
\hline
\rule{0mm}{4mm}
\rule{1mm}{0mm} & four-body terms($\eta$) \rule{10mm}{0mm} &
$\hat{h}^{\langle abcd\rangle}_{\eta}$ 
& ${\cal J}_\eta$\rule{4mm}{0mm} & process \rule{2mm}{0mm} \\
\hline
&ring &${\mathcal J}_{\text{ring}}(\;\bm{S}_{a}\cdot\bm{S}_{b})(\bm{S}_{c}\cdot\bm{S}_{d})$ & $40(c^4+c^2s^2+s^4)$ & $4b4s$ \\
&ring$^{\text{(d)}}$ &${\mathcal J}_{\text{ring}^{(d)}}\;
(\bm{S}_{a}\cdot\bm{S}_{c})(\bm{S}_{b}\cdot\bm{S}_{d})$& $-20(c^4+s^4)$ & $4b4s$ \\
& ring-KS & ${\mathcal J}_{\text{ring-KS}}\;
(\bm{n}_{ab}\cdot\bm{S}_{a})(\bm{n}_{bc}\cdot\bm{S}_{b})( \bm{n}_{cd}\cdot\bm{S}_{c})(\bm{n}_{da}\cdot\bm{S}_{d})$& $160s^4$ & $4b4s$  \\
& KS$\times$H &${\mathcal J}_{\text{KS}\times \text{H}}\;
(\bm{n}_{cd}\cdot\bm{S}_{a})(\bm{n}_{da}\cdot\bm{S}_{b})(\bm{S}_{c}\cdot\bm{S}_{d})$& $-80(c^2+2s^2)s^2$ & $4b4s$ \\
& KS$\times$H$^{\text{(d)}}$ &${\mathcal J}_{{\text{KS}\times\text{H}}^{\rm (d)}}\;
   (\bm{n}_{ab}\cdot(\bm{S}_{a}\times\bm{S}_{b}))(\bm{n}_{cd}\cdot\bm{S}_{c})(\bm{n}_{cd}\cdot\bm{S}_{d})$& $-80c^2s^2$ & $4b4s$ \\
& DM$\times$KS &${\mathcal J}_{{\text{DM}\times\text{KS}}}\;
   (\bm{n}_{ab}\cdot(\bm{S}_{a}\times\bm{S}_{b}))(\bm{n}_{cd}\cdot\bm{S}_{c})(\bm{n}_{cd}\cdot\bm{S}_{d})$& $-320cs^3$ & $4b4s$ \\
& DM$\times$H & ${\mathcal J}_{{\text{DM}\times\text{H}}}\;
  (\bm{n}_{ab}\cdot(\bm{S}_{a}\times\bm{S}_{b}))(\bm{S}_{c}\cdot\bm{S}_{d})$& $-160(c^2-s^2)cs$ & $4b4s$  \\
& DM$\times$DM &  ${\mathcal J}_{\rm DM \times DM}
  (\bm{n}_{ab}\cdot(\bm{S}_{a}\times\bm{S}_{b}))(\bm{n}_{cd}\cdot(\bm{S}_{c}\times\bm{S}_{d}))$& $-40c^2 s^2$ & $4b4s$  \\
& const & & $c^4+2c^2s^2-s^4$ & $4b4s$  \\
\hline
\rule{0mm}{4mm}\rule{1mm}{0mm} &two-body terms ($\eta$) \rule{10mm}{0mm} &
${\hat h}^{(ij)}_{\eta}$
& ${\cal J}_\eta$\rule{4mm}{0mm}  & process \rule{2mm}{0mm} \\
\hline
&H &${\mathcal J}_{\rm H} \;\bm{S}_{i}\cdot\bm{S}_{j}$& $-16(c^4-s^4)$ & $1b2s$  \\
&H &${\mathcal J}_{\rm H} \;\bm{S}_{i}\cdot\bm{S}_{j}$& $-4(c^4-s^4)$ & $4b4s$  \\
&H$_{\textrm{d}}$ &${\mathcal J}_{\rm H^{(\textrm{d})}} \;\bm{S}_{i}\cdot\bm{S}_{k}$& $-2(c^2+s^2)^2$ & $4b4s$ \\
&H$_{\textrm{d}}$ &${\mathcal J}_{\rm H^{(\textrm{d})}} \;\bm{S}_{i}\cdot\bm{S}_{k}$& $4(c^4-6c^2s^2+s^4)$ & $2b3s$ \\
&KS &${\mathcal J}_{\rm KS} \; ({\bm n}_{ij}\cdot\bm{S}_{i})({\bm n}_{ij}\cdot\bm{S}_{j})$& $-32(c^2+s^2)s^2$  &$1b2s$ \\
&KS&${\mathcal J}_{\rm KS} \; ({\bm n}_{ij}\cdot\bm{S}_{i})({\bm n}_{ij}\cdot\bm{S}_{j})$& $8(3c^2-s^2)s^2$   &$4b4s$\\
&KS$^{(\textrm{d})}$ &${\mathcal J}_{\rm KS^{(\textrm{d})}} \; ({\bm n}_{ij}\cdot\bm{S}_{i})({\bm n}_{jk}\cdot\bm{S}_{k})$& $16c^2s^2$  & $4b4s$ \\
&KS$^{(\textrm{d})}$ &${\mathcal J}_{\rm KS^{(\textrm{d})}} \; ({\bm n}_{ij}\cdot\bm{S}_{i})({\bm n}_{jk}\cdot\bm{S}_{k})$& $32c^2s^2$  & $2b3s$ \\
&DM &${\mathcal J}_{\rm DM} \;{\bm n}_{ij} \cdot (\bm{S}_{i}\times\bm{S}_{b})$& $32cs$ & $1b2s$ \\
&DM &${\mathcal J}_{\rm DM} \;{\bm n}_{ij} \cdot (\bm{S}_{i}\times\bm{S}_{b})$& $8cs$ & $4b4s$ \\
&const & & $16$ & $1b2s$ \\
&const$^{\textrm{(d)}}$ & & $-4$ & $2b3s$ \\
\hline
\end{tabular}
\label{tab1}
\end{table*}

\begin{table*}
\caption{
Constituents of the fourth order Hamiltonian ${\cal H}^{(4)}$ for type-N with broken-inversion symmetry. 
$c$ and $s$ included in the coefficients ${\cal J}_\eta$. 
The first seven rows of four-body terms separated from by line are those that appear also in type-C, 
whereas the next eight terms are present only in type-N. 
The first eight rows of the two-body term appear in type-C, 
and the rest appears only in type-N. 
The energy unit is ${t_{\rm eff}^4}{U^3}$, see Eq.(\ref{eq:4th}).
}
\begin{tabular}{lllrc}
\hline
\rule{0mm}{4mm}
\rule{1mm}{0mm} & four-body terms($\eta$)  \rule{4mm}{0mm} & 
$\hat{h}^{\langle abcd\rangle}_{\eta}$
&${\cal J}_\eta$ \rule{4mm}{0mm} & process \rule{2mm}{0mm} \\
\hline
&ring &$ {\mathcal J}_{\text{ring}}(\;\bm{S}_{a}\cdot\bm{S}_{b})(\bm{S}_{c}\cdot\bm{S}_{d})$& $40(c^4+2c^2s^2-s^4)$ & $4b4s$ \\
&ring$^{\text{(d)}}$ &${\mathcal J}_{\text{ring}}^{(d)}\;
(\bm{S}_{a}\cdot\bm{S}_{c})(\bm{S}_{b}\cdot\bm{S}_{d})$& $-20(c^4+2c^2s^2-s^4)$ & $4b4s$ \\
& ring-KS &$ {\mathcal J}_{\text{ring-KS}}\;
(\bm{n}_{ab}\cdot\bm{S}_{a})(\bm{n}_{bc}\cdot\bm{S}_{b})( \bm{n}_{cd}\cdot\bm{S}_{c})(\bm{n}_{da}\cdot\bm{S}_{d})$& $160s^4$ & $4b4s$  \\
& KS$\times$H$^{\textrm{(d)}}_1$ &${\mathcal J}_{{\text{KS}\times\text{H}}_1^{(d)}}\;
   (\bm{n}_{ab}\cdot(\bm{S}_{a}\times\bm{S}_{b}))(\bm{n}_{cd}\cdot\bm{S}_{c})(\bm{n}_{cd}\cdot\bm{S}_{d})$& $80c^2s^2$ & $4b4s$ \\
& DM$\times$H &${\mathcal J}_{{\text{DM}\times\text{H}}}\;
  (\bm{n}_{ab}\cdot(\bm{S}_{a}\times\bm{S}_{b}))(\bm{S}_{c}\cdot\bm{S}_{d})$& $160c^3s$ & $4b4s$ \\
& DM$\times$DM &${\mathcal J}_{\rm DM \times DM}
  (\bm{n}_{ab}\cdot(\bm{S}_{a}\times\bm{S}_{b}))(\bm{n}_{cd}\cdot(\bm{S}_{c}\times\bm{S}_{d}))$& $40(c^2s^2-s^4)$ & $4b4s$  \\
& const &$ $& $c^4+2c^2s^2-s^4$ & $4b4s$ \\
\hline
& KS$\times$H$_1$ &\begin{tabular}{l}${\mathcal J}_{\rm KS\times H_1}
\big((\bm{n}_{da}\cdot\bm{S}_{a})(\bm{n}_{ab}\cdot\bm{S}_{b})$\\
$\hspace{15mm}+(\bm{n}_{ab}\cdot\bm{S}_{a})(\bm{n}_{bc}\cdot\bm{S}_{b})\big)(\bm{S}_{c}\cdot\bm{S}_{d})
\big)$\end{tabular}& $-160c^2s^2$ & $4b4s$ \\
& KS$\times$H$_2$ &\begin{tabular}{l}${\mathcal J}_{\rm KS\times H_2}
\big((\bm{n}_{ab}\cdot\bm{S}_{a})(\bm{n}_{cd}\cdot\bm{S}_{b})$
$+(\bm{n}_{cd}\cdot\bm{S}_{a})(\bm{n}_{ab}\cdot\bm{S}_{b})\big)(\bm{S}_{c}\cdot\bm{S}_{d})\big)$\end{tabular}& $80(c^2s^2-s^4)$ & $4b4s$ \\
& KS$\times$H$^{\textrm{(d)}}_2$ &\begin{tabular}{c}${\mathcal J}_{{\text{KS}\times\text{H}}_2^{(d)}}\big((\bm{n}_{ab}\cdot\bm{S}_{a})(\bm{n}_{cd}\cdot\bm{S}_{b})$
$+(\bm{n}_{cd}\cdot\bm{S}_{a})(\bm{n}_{ab}\cdot\bm{S}_{b})\big)(\bm{S}_{c}\cdot\bm{S}_{d})\big)$\end{tabular}& $-80(c^2-s^2)s^2$ & $4b4s$ \\
& DM$\times$KS$^{\textrm{(d)}}$ &\begin{tabular}{c}${\mathcal J}_{{\text{DM}\times\text{KS}}^{(d)}}\!\big((\bm{n}_{ab}\cdot(\bm{S}_{a}\times\bm{S}_{c}))(\bm{n}_{bc}\cdot\bm{S}_{b})(\bm{n}_{da}\cdot\bm{S}_{d})$\\
   $+(\bm{n}_{bc}\cdot(\bm{S}_{a}\times\bm{S}_{c}))(\bm{n}_{ab}\cdot\bm{S}_{b})(\bm{n}_{cd}\cdot\bm{S}_{d})\big)$\end{tabular}& $-160cs^3$ & $4b4s$  \\
& DM$\times$DM$_2$ &${\mathcal J}_{\rm DM \times DM_2}
 (\bm{n}_{bc}\cdot(\bm{S}_{a}\times\bm{S}_{b}))(\bm{n}_{da}\cdot(\bm{S}_{c}\times\bm{S}_{d}))$& $-40(c^2s^2-s^4)$ & $4b4s$  \\
& DM$\times$Ising$^{\textrm{(d)}}$ &$\mathcal{J}_{\text{DM}\times\text{Ising}^{(d)}}(\bm{n}_{cd}+\bm{n}_{da})\cdot(\bm{S}_{a}\times\bm{S}_{c})(S_{b}^z\cdot S_{d}^z)$& $-80cs^3$ & $4b4s$ \\
& DM$\times$H$^{\textrm{(d)}}$ &${\mathcal J}_{\rm DM\times H^{(d)}}
(\bm{n}_{ab}\cdot(\bm{S}_{a}\times\bm{S}_{c}))(\bm{S}_{b}\cdot \bm{S}_{d})$& $160c^3s$ & $4b4s$ \\
& ring-$\Gamma$ &\begin{tabular}{l}$\mathcal{J}_{\text{ring-}\Gamma}\big((\bm{n}_{ab}\cdot\bm{S}_a)S_b^{z}+S_a^{z}(\bm{n}_{ab}\cdot\bm{S}_b)\big)
$
$(\bm{n}_{bc}\cdot\bm{S}_c)(\bm{n}_{da}\cdot\bm{S}_d)$\end{tabular}& $-160cs^3$ & $4b4s$ \\
\hline
\rule{0mm}{4mm}\rule{1mm}{0mm} &two-body terms ($\eta$) \rule{10mm}{0mm} &  
${\hat h}^{(ij)}_{\eta}$
&${\cal J}_\eta$ \rule{4mm}{0mm}  & process \rule{2mm}{0mm} \\
\hline
&H &${\mathcal J}_{\rm H} \;\bm{S}_{i}\cdot\bm{S}_{j}$& $-4(c^4+s^4)$ & $4b4s$  \\
&H &${\mathcal J}_{\rm H} \;\bm{S}_{i}\cdot\bm{S}_{j} $& $-16(c^4-s^4)$ & $1b2s$ \\
&H$^{\textrm{(d)}}$ &${\mathcal J}_{\rm H^{(\textrm{d})}} \;\bm{S}_{i}\cdot\bm{S}_{k}$& $-2(c^4-2c^2s^2-s^4)$ & $4b4s$  \\
&H$^{\textrm{(d)}}$ &${\mathcal J}_{\rm H^{(\textrm{d})}} \;\bm{S}_{i}\cdot\bm{S}_{k}$& $4(c^4-2c^2s^2+s^4)$ & $2b3s$  \\
&DM &${\mathcal J}_{\rm DM} \;{\bm n}_{ij} \cdot (\bm{S}_{i}\times\bm{S}_{b})$& $-8c^3s$ & $4b4s$ \\
&DM &${\mathcal J}_{\rm DM} \;{\bm n}_{ij} \cdot (\bm{S}_{i}\times\bm{S}_{b})$& $32cs$ & $1b2s$ \\
&const &$ $& $16$ & $1b2s$ \\
&const$^{\textrm{(d)}}$ &$ $& $-4$ & $2b3s$ \\
\hline
&KS$_1$ &${\mathcal J}_{\rm KS_1}
(\bm{n}_{ij}\cdot\bm{S}_{i})(\bm{n}_{ji}\cdot\bm{S}_{j})$& $8(c^2s^2-s^4)$ & $4b4s$ \\
&KS$_2$ &$ \mathcal{J}_{\rm KS_2}\big((\bm{n}_{ij}\cdot\bm{S}_{i})(\bm{n}_{jk}\cdot\bm{S}_{j})
 -(\bm{n}_{kj}\cdot\bm{S}_{i})(\bm{n}_{ji}\cdot\bm{S}_{j})\big)$& $-8c^2s^2$  & $4b4s$\\
&KS$_1$ &${\mathcal J}_{\rm KS_1}
(\bm{n}_{ij}\cdot\bm{S}_{i})(\bm{n}_{ji}\cdot\bm{S}_{j})$& $32(c^2+s^2)s^2$ & $1b2s$ \\
&KS$_1^\textrm{(d)}$ &$\mathcal{J}_{\rm KS_1^{\textrm{(d)}}}\big((\bm{n}_{ij}\cdot\bm{S}_{i})(\bm{n}_{jk}\cdot\bm{S}_{k})+
 (\bm{n}_{kj}\cdot\bm{S}_{i})(\bm{n}_{ji}\cdot\bm{S}_{k})\big)$& $4c^2s^2$  & $4b4s$ \\
&KS$_2^\textrm{(d)}$ &$\mathcal{J}_{\rm KS_2^{\textrm{(d)}}}\big((\bm{n}_{ij}\cdot\bm{S}_{i})(\bm{n}_{ji}\cdot\bm{S}_{k})
 +(\bm{n}_{jk}\cdot\bm{S}_{i})(\bm{n}_{kj}\cdot\bm{S}_{k})\big)$& $8(c^2s^2+s^4)$  & $4b4s$ \\
&KS$_1^\textrm{(d)}$ &$\mathcal{J}_{\rm KS_1^{\textrm{(d)}}}\big((\bm{n}_{ij}\cdot\bm{S}_{i})(\bm{n}_{jk}\cdot\bm{S}_{k})+
 (\bm{n}_{kj}\cdot\bm{S}_{i})(\bm{n}_{ji}\cdot\bm{S}_{k})\big)$& $16c^2s^2$ & $2b3s$ \\
&KS$_2^\textrm{(d)}$ &$\mathcal{J}_{\rm KS_2^{\textrm{(d)}}}\big((\bm{n}_{ij}\cdot\bm{S}_{i})(\bm{n}_{ji}\cdot\bm{S}_{k})
 +(\bm{n}_{jk}\cdot\bm{S}_{i})(\bm{n}_{kj}\cdot\bm{S}_{k})\big)$& $16c^2s^2$ & $2b3s$ \\
&DM$^{\textrm{(d)}}$ &${\mathcal J}_{\rm DM^{\text(d)}}
\big((\bm{n}_{ij}+\bm{n}_{jk})\cdot(\bm{S}_{i}\times\bm{S}_k))$& $-8c^3s$ & $4b4s$ \\
&DM$^{\textrm{(d)}}$ &${\mathcal J}_{\rm DM^{\text(d)}}
\big((\bm{n}_{ij}+\bm{n}_{jk})\cdot(\bm{S}_{i}\times\bm{S}_k))$& $-16c^3s$ & $2b3s$ \\
&$\Gamma$ &${\mathcal J}_{\Gamma}
\big((\bm{n}_{ij}\cdot\bm{S}_i)S_j^{z}+S_i^{z}(\bm{n}_{ij}\cdot\bm{S}_j)\big)$& $8cs^3$ & $4b4s$ \\
\hline
\end{tabular}
\label{tab2}
\end{table*}

\subsection{Second order terms}
We first briefly introduce the second-order terms known from the previous literature,
which become the building blocks of the fourth-order terms we derive shortly.
The second-order Hamiltonian consists of exchange interactions between neighboring spins
given as
\begin{align}
  \mathcal{H}^{(2)}&=\sum_{\langle i,j\rangle}
  \big(J\bm{S}_i\cdot\bm{S}_j + \bm{D}_{ij}\cdot\bm{S}_i\times\bm{S}_j
\notag\\
& \rule{10mm}{0mm} +K(\bm{n}_{ij} \cdot\bm{S}_i)(\bm{n}_{ij}\cdot\bm{S}_j) + {\cal C}^{(2)} \big).
\label{eq:h2}
\end{align}
Here, $J=(4t_{\rm eff}^2/U) \cos\theta$ is the Heisenberg term,
$\bm{D}_{ij}=\bm{n}_{ij} (4t_{\rm eff}^2/U) \sin\theta$ and $K=\sqrt{J^2+D^2}-J$
are the spin anisotropic terms denoted as
Dzyaloshinskii-Moriya (DM) \cite{Moriya1960,Dzyaloshinsky1958}
and Kaplan-Shekhtman-Aharony-Entin-Wohlman (KSAEW) terms\cite{Kaplan1983,Shekhtman1992,Shekhtman1993},
respectively.
Importantly, the DM vector $\bm D_{ij}$ and the KSAEW spin anisotropy axis vector are both parallel to $\bm n_{ij}$.
The DM interaction twists the spins and makes them non-collinear
within the plane perpendicular to $\bm{D}_{ij}$.
The KSAEW terms act as the antiferromagnetic Ising interaction along $\bm{n}_{ij}$, which bring an anisotropy.
The constant term, $-t_{\rm eff}^2/U$, is shown explicitly for later convenience.

\subsection{Third order terms}
For the third order processes, we consider three-site unit $(a b c)$ forming a closed triangular loop. 
This process needs to be considered when there are such nonzero hopping paths on the lattice; 
On a square lattice with only nearest neighbor hopping term, this process is trivially absent. 
\ch{Additionally}, it is zero even for the triangular lattice, which can be explained as follows;  
Third-order processes are classified into clockwise and counterclockwise processes along the triangular loop. 
Figure~\ref{f1}(c) shows the example of a pair of processes that contribute to $\mathcal{H}^{(3)}_{\rm eff}$.
The two processes have the opposite signs and cancel out.
Such cancellation occurs for all pairs and we find exactly $\mathcal{H}^{(3)}_{\rm eff}=0$.

\subsection{Fourth order terms}
For the fourth-order process, we consider several types of paths around the plaquette,
labeled according to the number of bonds, $(1b, 2b, 4b)$, and sites, $(2s, 3s, 4s)$,
that participate in the process.
We show in Fig.~\ref{f1}(d) four different types of processes.
Among them, the $(2b4s)$ process cancels out\cite{Tanaka2018,Calzado2004}.
\par
As mentioned, the fourth-order Hamiltonian consists of both four-body and two-body terms,
which is given as a sum of terms that operate on the \ch{plaquette labeled $\gamma$} as
\begin{align}
\label{eq:4th}
 \mathcal{H}^{(4)}=& \;\frac{t_{\rm eff}^4}{U^3} \;\bigg(
\sum_{\gamma} \sum_{\langle abcd \rangle} \hat{h}^{\langle abcd \rangle}_\gamma  +
 \sum_{\gamma'} \sum_{(ij)} \hat{h}^{(ij)}_{\gamma'}  \bigg)
\end{align}
Here, for each plaquette, we take the sum cyclically as
$\langle abcd\rangle= (1234), (2341), (3412), (4123)$ over four spins labeled as 1,2,3,4
around the plaquette.
For the two-body term, we take the summation only once for each pair without duplication;
we consider all possible pairs of spins $(ij)$, e.g. (12),(23),(34),(41)
for nearest neighbor. 
For the next-nearest neighbor, we have (13), (24), or (1$x$), (2$x$),$\cdots$, with 
either $i$ or $j$ belonging to the plaquette and $x$ being its next-nearest site outside. 
For those inside the plaquette, e.g. $(ik)=(13)$, 
we have two choices of $j$ but we define $(i \to j \to k)$ as being clockwise, 
where the index $j$ appears explicitly in the Hamiltonian. 
The constant terms for each pair or plaquette are also included for completeness, 
whose summations are taken in the same manner as other terms. 
\par
In the following, we present the explicit form of $\hat{h}^{\langle abcd \rangle}_\gamma$ and 
$\hat{h}^{(ij)}_\gamma$ for a \ch{plaquette labeled $\gamma$}. 
As shown in Eq.(\ref{eq:4th}), the energy unit of these terms is $t^4_{\rm eff}/U^3$, 
and $\mathcal{H}^{(4)}$ is obtained by taking the sum of all possible choices of $\gamma$ 
on the given lattice. 
The example of extracting $\gamma$ from the square lattice with one diagonal bond is shown 
in Fig.~\ref{f1}(b). 
How these fourth-order terms are derived systematically depending on a series of hopping processes
\ch{is} explained in Appendix \ref{app:exchange}. 
\begin{figure*}[t]
  \includegraphics[width=1.0\textwidth]{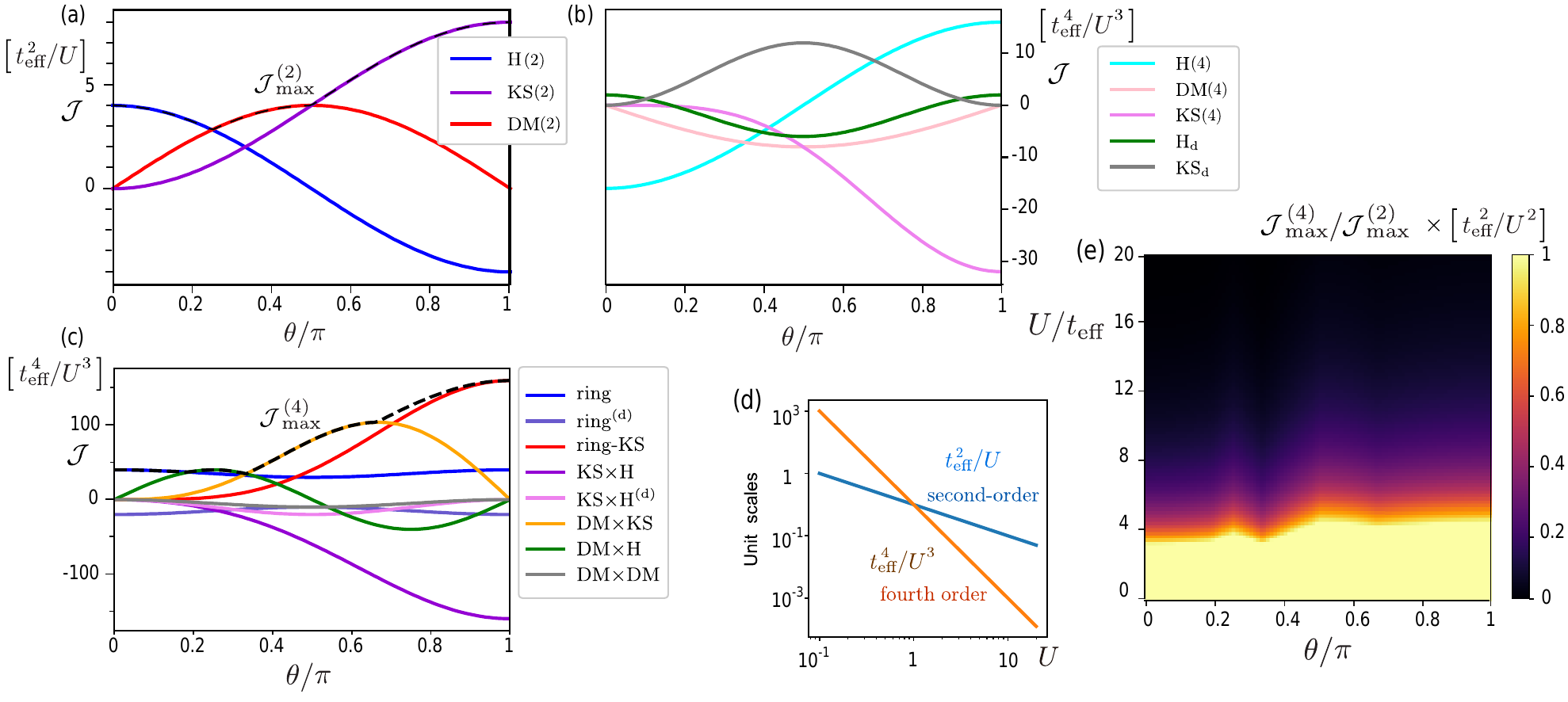}
  \caption{Exchange coupling constants for type-C SOC in Eq.(\ref{eq:4th-A}) and Table \ref{tab1}.
  Coupling constants of  (a,b) the two-body and (c) four-body terms, 
  where we take the energy unit as $[t_{\rm eff}^2/U]$ and $[t_{\rm eff}^3/U^2]$ 
  for the second order (a) and fourth order terms (b,c), respectively, and set $U/t_{\rm eff}=10$. 
  For the second order term in panel (a), we plot the values, 
  ${\mathcal J}= J, D$, and $K$ divided by $t_{\rm eff}^2/U$. 
  (d) Unit energy scales, $t_{\rm eff}^2/U$ (second order) and $t_{\rm eff}^3/U^2$ (fourth order),
      as functions of $U/t_{\rm eff}$. 
  (e) Density plot of the maximum coupling constant among the fourth order terms ${\mathcal J}^{(4)}_{\rm max}$
     against the second order one ${\mathcal J}^{(2)}_{\rm max}$ on the plane of $\theta$ and $U$, where we set $t_{\rm eff}=1$. 
     The variations of ${\mathcal J}^{(2)}_{\rm max}$ and ${\mathcal J}^{(4)}_{\rm max}$ as functions of $\theta$ are shown explicitly 
    in panels (a,c). }
  \label{f-Comp}
\end{figure*}
\begin{figure}[tbp]
  \includegraphics[width=0.45\textwidth]{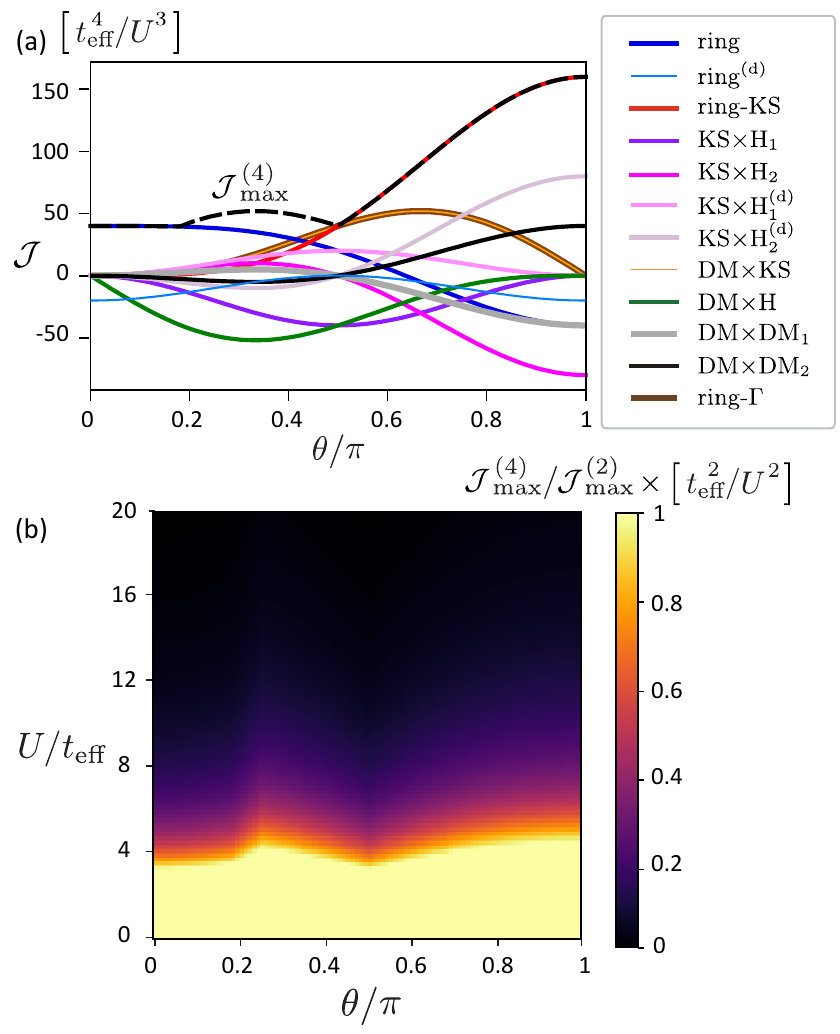}
  \caption{Exchange coupling constants for type-N SOC coupling in Table \ref{tab2}.
  (a) Coupling constants of the fourth order terms
  in a unit of $[t_{\rm eff}^3/U^4]$, where we set $U/t_{\rm eff}=10$.
  (b) Density plot of the maximum coupling constant among the fourth order terms
     against the second order one on the plane of $\theta$ and $U$, where we set $t_{\rm eff}=1$.
  }
  \label{f-Rashba}
\end{figure}
%
\subsubsection{Type-C with inversion symmetry}
In Table~\ref{tab1}, we list all the terms 
that appear in the fourth order process with uniform $\bm n_{ij}=\bm n$ perpendicular to the plane.
Here, $\hat{h}^{\langle abcd \rangle}_\eta$ ($\eta$ is the label of the term) 
is given as a combination of 
spin operators and its coefficients ${\mathcal J}_{\eta}(\theta)$. 
For comprehensiveness of the notation, we show explicitly the first two rows in Table~\ref{tab1} 
from among seven different types of four-body terms as 
\begin{align}
\label{eq:4th-A}
\hat{h}^{\langle abcd\rangle}_{\text{ring}}
 \rule{1mm}{0mm} &= {\mathcal J}_{\text{ring}}\;
(\bm{S}_{a}\cdot\bm{S}_{b})(\bm{S}_{c}\cdot\bm{S}_{d}),
\notag\\
\hat{h}^{\langle abcd\rangle}_{\rm ring^{\text{(d)}}}
 \rule{1mm}{0mm} &= {\mathcal J}_{\text{ring}}^{(d)}\;
(\bm{S}_{a}\cdot\bm{S}_{c})(\bm{S}_{b}\cdot\bm{S}_{d}), \cdots. 
\end{align} 
and 
\begin{align}
\label{eq:2ndex}
{\hat h}^{(ij)}_{\rm H}&={\mathcal J}_{\rm H} \;\bm{S}_{i}\cdot\bm{S}_{j}, \notag\\
{\hat h}^{(ik)}_{\rm KS^{(\textrm{d})}} &=
{\mathcal J}_{\rm KS^{(\textrm{d})}} \; ({\bm n}_{ij}\cdot\bm{S}_{i})({\bm n}_{jk}\cdot\bm{S}_{k}), \cdots, 
\end{align}
for the two-body terms. 
As mentioned, ${\hat h}^{(ik)}_{\eta}$ is the next-nearest neighbor interactions between $\bm S_i$ and 
$\bm S_k$ with $\bm S_j$ in between, where $(i \to j \to k)$ is taken clockwise.
\par
Notice that the types of exchange interactions do not have clear correspondence with the 
perturbation processes. 
For example, one of the fourth order processes that has initial and final state, 
$(S_a^z, S_b^z, S_c^z, S_d^z)=(\uparrow,\downarrow,\downarrow,\uparrow)$, 
and $(\downarrow,\uparrow,\uparrow,\downarrow)$  
contributes to both ring and ring$^{(d)}$ terms in Eq.(\ref{eq:2ndex}). 
Therefore, to classify numerous terms, 
we rely on the $\theta$-dependence that appear in ${\cal J}_\eta$ 
(see Appendix A for details).  
\par
We have for the two-body terms, five different types, 
while each of them appears twice in Table I, classified according to the processes, e.g. ($1b2s$), they belong to. 
These terms basically reduce to the same form as Eq.(\ref{eq:h2}), 
although the amplitudes at large $U$ differ by orders of magnitude, 
$t_{\rm eff}^2/U$ and $t_{\rm eff}^4/U^3$, for the second and fourth order. 
The difference from the second order is that there are interactions between next nearest neighbor spins
due to $2b3s$ process in both the diagonal and bond directions (see Fig.~\ref{f1}(d)).
These next-nearest neighbor two-body terms have superscript (d) as listed in Table I. 
\par
When $\theta=0$,
the four-body terms in this Hamiltonian reduces to
the ring exchange term obtained for Hubbard model \cite{Takahashi1977},
\begin{align}
\mathcal{H}_{\rm ring}^{(4)}
&= 80\frac{t_{\rm eff}^4}{U^3} \big( (\bm S_1\cdot \bm S_2)(\bm S_3 \cdot \bm S_4)+(\bm S_1\cdot \bm S_4)(\bm S_2 \cdot \bm S_3)
\notag \\s
& -(\bm S_1\cdot \bm S_3)(\bm S_2 \cdot \bm S_4)\big).
\end{align}
\subsubsection{Type-N with broken inversion symmetry}
We now consider the Rashba-type SOC.
Since $\bm n_{ij}$ points in-plane and rotates counterclockwise when the electron
travels clockwise around the plaquette, different types of spin components may mix and
give further variety in the types of spin exchange interactions,
as listed in Table~\ref{tab2}, 
where the first seven rows are common in their operator form to type-C ones 
and the lower eight terms are found only in type-N.
\par
Among the two-body terms, we also find anisotropic spin exchange terms like \ch{the} $\Gamma$ term or KS$_{1}$ or KS$_2$ terms 
that did not appear in type-C, which are shown in the bottom 10 rows of Table II. 
From the form of coefficients given in Table~\ref{tab2}, we find the two-body terms except
DM and KS$_1^{\text{(d)}}$ are \ch{smaller} by one order of magnitude from the four-body terms
and are considered irrelevant.
\par
\section{Basic properties of the effective Hamiltonian}
\subsection{Effect of SOC on the exchange coupling constants}
We now examine the basic nature of the obtained effective Hamiltonian.
Figures~\ref{f-Comp}(a)-(c) show the $\theta$ dependence of the coupling constants 
of second order terms (Eq.(\ref{eq:h2})) ${\cal J}=J, D, K$, 
and fourth order terms for type-C Hamiltonian Table \ref{tab1},
normalized by $t_{\rm eff}^2/U$ and $t_{\rm eff}^4/U^3$, respectively, 
for fixed $U/t_{\rm eff}=10$. 
\par
In Fig.~\ref{f-Comp}(a) and \ref{f-Comp}(b), we plot the two-body terms ${\cal J}=J, D, K$, 
that appear in ${\cal H}^{(2)}$ 
and ${\cal H}^{(4)}$, respectively;
the Heisenberg interaction transforms from the antiferromagnetic one
to the ferromagnetic one at $\theta=\pi/2$.
The DM interaction is the most dominant when $\theta=\pi/2$,
whereas, the KSAEW term increases with $\theta$ and takes the maximum at $\theta=\pi$.
Therefore, there is an overall tendency that $\theta\sim 0$ is the antiferromagnet,
$\theta\sim \pi/2$ twists the spins to be noncollinear with a strong tendency to form a long wavelength
swirling structure, and $\theta=\pi$ is the Ising ferromagnet.
The contributions from the second and fourth order show similar $\theta$-dependence,
and the amplitude of the latter is \ch{$\sim t_{\rm eff}^2/U^2=0.01$} of the former, indicating that
the H, DM, and KS terms are equally amplified by \ch{$\sim 1.01$}.
by the inclusion of the fourth-order terms.
\par
Figure~\ref{f-Comp}(d) shows how the unit energy scale, 
$t_{\rm eff}^2/U$ and $t_{\rm eff}^4/U^3$, of the second and fourth order evolves
with $U$ when setting $t_{\rm eff}=1$. 
With this, one can access the absolute values of ${\cal J}$'s 
in panels (a) and (b,c) by multiplying $t_{\rm eff}^2/U$ and $t_{\rm eff}^4/U^3$, respectively, for the given value of $U$.
With this information in mind, we evaluate the relative intensity 
of the largest fourth-order coupling constants ${\cal J}^{(4)}_{\rm max}$ 
against the second-order one ${\cal J}^{(2)}_{\rm max}$ in Fig.~\ref{f-Comp}(c), 
where we multiply the relative amplitudes of their units, $(t_{\rm eff}^2/U^2)$. 
Here, ${\cal J}^{(2)}_{\rm max}$ is the largest among $J, D$, and $K$ divided by $(t_{\rm eff}^2/U)$, 
shown in broken line in Fig.~\ref{f-Comp}(a).  
Interestingly, when $U/t_{\rm eff}\lesssim 10$, the fourth order term develops and 
reaches almost half the second order ones at $U/t_{\rm eff}\sim 8$. 
This tendency is kept for all values of $\theta$. 
In particular, when $\theta \gtrsim 0.5\pi$ where the spin-dependent hopping term $\lambda$ overwhelms
the standard hopping term $t$, the anisotropic exchange interactions
like ring-KS, KS$\times$H, and DM$\times$KS become dominant and may impact the nature of magnetism.
The Heisenberg-related terms like KS$\times$H are small enough to be discarded.
This situation can be quite often observed in 4$d$ or 5$d$ compounds with substantial atomic SOC
when the crystal field symmetry is lowered.
\par
The same analysis for type-N is shown in Fig.~\ref{f-Rashba}(a),
where we have the Rashba SOC with broken inversion symmetry.
Several differences from type-C are observed.
First of all, the types of interactions have more variety,
and most of them have comparable values ranging from $-0.05$ to $0.05$,
except for the ring-KS term and KS$\times$H$_2$ that are enhanced toward $\theta\sim \pi$.
Figure~\ref{f-Rashba}(b) shows the relative intensity of the fourth-order coupling constants compared
to the second-order ones, which is similar to type-N.
A wide variety of terms come from the enhanced Ising anisotropy in-plane that
projects the spins to the bond-dependent $\bm n_{ij}$, which is included in the 
terms having KS. 
These terms may overall help to develop a vortex spin structure on a square unit,
actually observed at $\theta\sim \pi$ in the square lattice ground state\cite{Kawano2023}.
Among these interactions, Fig.~\ref{f-Rashba}(a) tell us that 
DM$\times \rm H$ is important in the range of $\theta/\pi\sim 0.2-0.5$, 
consistent with the enhancement of second order DM at $\theta/\pi\sim 0.5$. 
At $\theta\sim\pi$ the KSAEW-related terms \ch{are dominant}. 
Ring-KS is \ch{particularly} enhanced, and the second largest ones 
are KS$\times \rm H_2$ and KS$\times \rm{H^{(d)}_2}$. 
What kind of spin configuration each term favors \ch{on an isolated square unit}
is discussed in Appendix \ref{app:groundstate} and Fig.~\ref{fGS}.
\par
\begin{figure}[tbp]
  \includegraphics[width=0.5\textwidth]{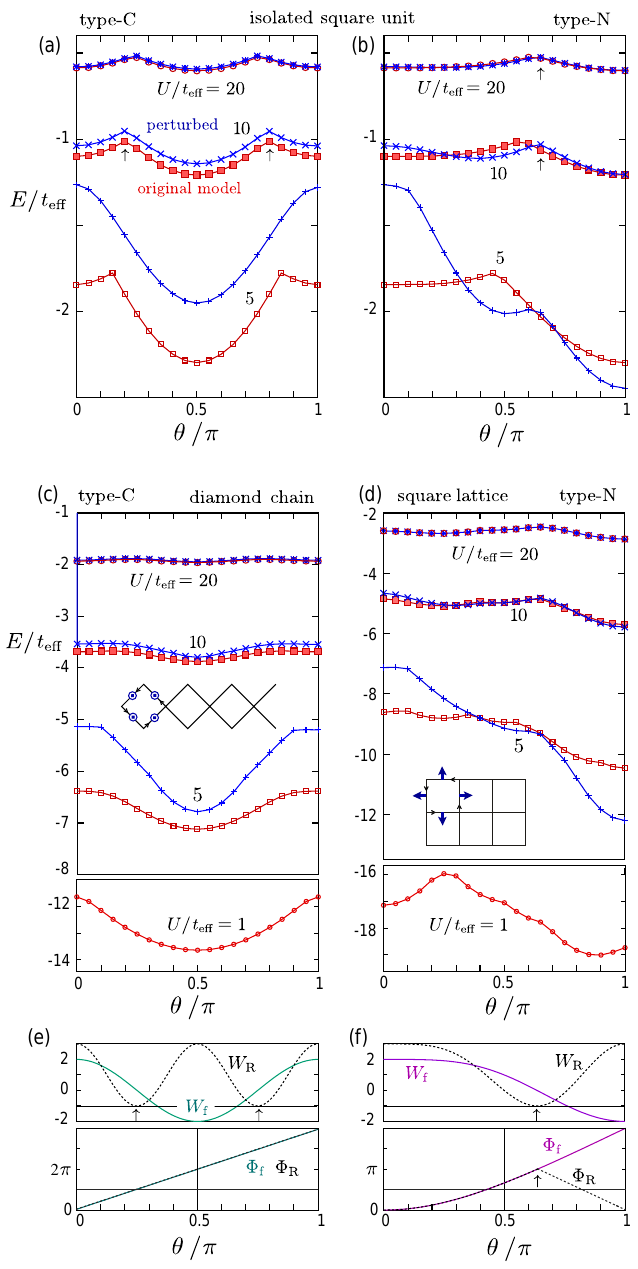}
  \caption{Energy eigenvalues $E$ compared between the effective spin Hamiltonian and the original Hamiltonian for
(a,b) the isolated square unit with type-C and type-N SOC,
and those of $N=12$ lattices, (c) diamond chain for type-C and (d) square lattice for type-N with periodic boundary conditions,
given for fixed $U/t_{\rm eff}=20,10,5$ (and $1$ for (c) and (d)) as functions of $\theta$,
where $\theta=0$ and $\pi$ correspond to the zero and full SOC hopping term, respectively.
(e,f) Wilson loop defined around the plaquette for fermions $W_{\text{f}}$ and spins $W_{\text{R}}$ and the corresponding
rotation angle $\Phi_{\text{f}}$ and $\Phi_{\text{f}}$ as functions of $\theta$. See Eqs.(\ref{eq:wlf}, \ref{eq:wlr}).
}
  \label{f4}
\end{figure}
\begin{figure*}[tbp]
  \includegraphics[width=1\textwidth]{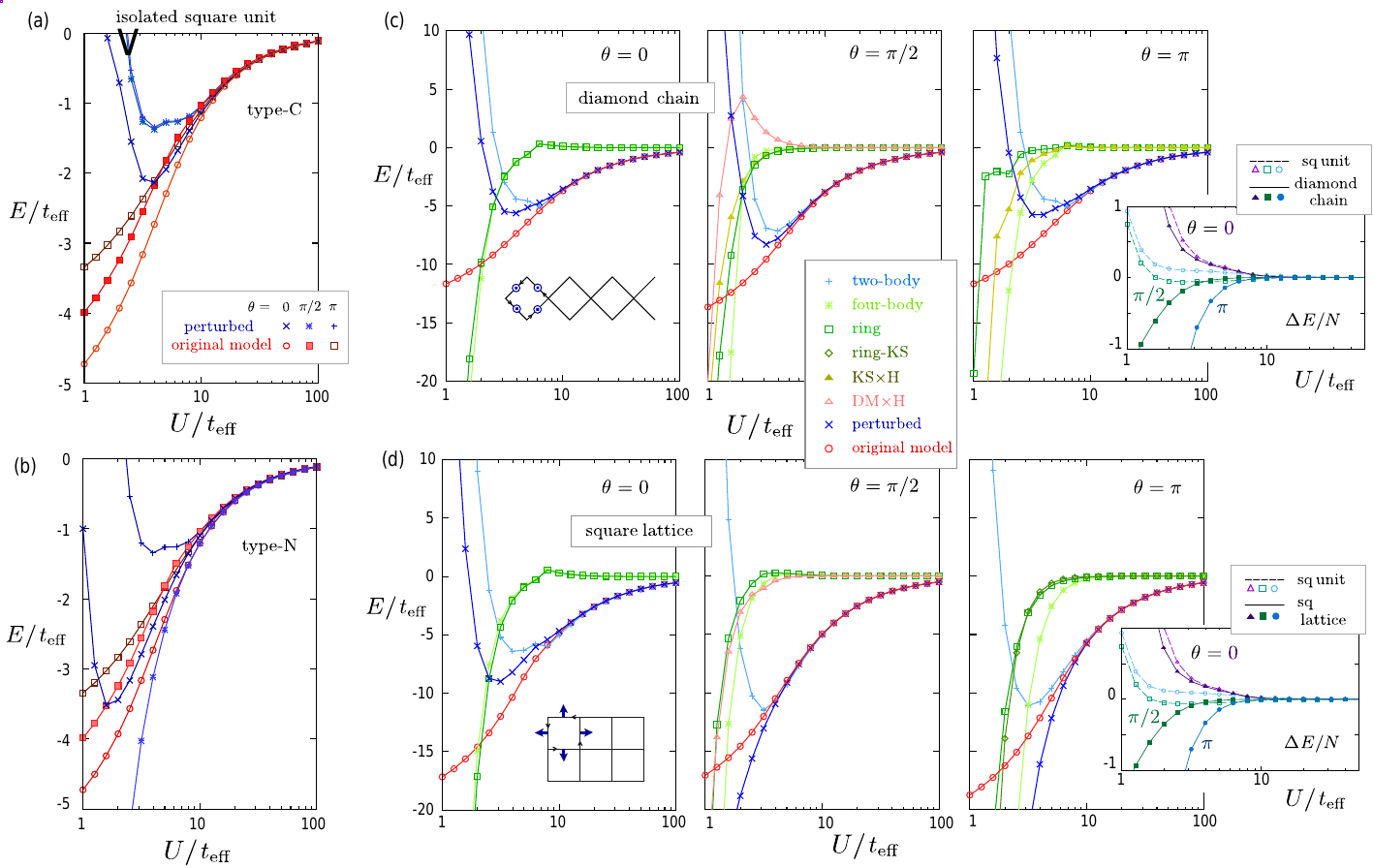}
  \caption{Energy eigenvalues $E$ compared between the effective spin Hamiltonian and the original Hamiltonian for
(a,b) the isolated square unit with type-C and type-N SOC, and those of $N=12$ lattices, (c) diamond chain for type-C and (d) square lattice for type-N with periodic boundary conditions, 
given for fixed $\theta=0,\pi/2,\pi$ as functions of $U/t_{\rm eff}$.
In panels (c,d) the energies of two-body and four-body terms and the representative spin exchange couplings of fourth orders
are shown together. 
The insets on the right show the energy difference per site, $\Delta E/N$, in unit of $t_{\rm eff}=1$ between 
the spin Hamiltonian and the original Hamiltonian 
for the isolated square units, diamond chain, and square lattice. }
  \label{f5}
\end{figure*}
\subsection{Comparison of the perturbed and original Hamiltonian}
We now compare the energies of the original Hamiltonian and
the perturbed effective spin Hamiltonian which includes all the terms up to fourth order.
Figures~\ref{f4}(a) and \ref{f4}(b) show the energy eigenvalue $E$ of these Hamiltonians
on an \ch{isolated} square unit $N=4$ with type-C and type-N SOC, respectively.
The energy of the spin model shows good agreement with that of the original model
at $U/t_{\rm eff}=10$, and almost perfectly coincides at $U/t_{\rm eff}=20$.
The profiles of the energy as functions of $\theta$ are 
symmetric about $\theta=\pi/2$ for type-C and not for type-N,
and there is a kink at $\theta/\pi=0.2, 0.8$ in type-C and
at around $\theta/\pi\sim 0.7$ in type-N. 
\par
To see how these tendencies may be sustained for larger system sizes,
we perform the exact diagonalization of $N=12$ cluster Hamiltonian with periodic boundaries,
where for type-C we choose the diamond chain and for type-N the square lattice,
shown inside the panels of Figs.~\ref{f4}(c) and \ref{f4}(d).
We chose the one-dimensional chain for type-C because the SOC with uniform
$\bm n\parallel\bm e_z$ is only compatible with corner-sharing lattices.
Although the kinks are smeared, the overall tendencies are well-preserved.
It is notable that the energy kink in type-N at $U/t_{\rm eff}=20$ is visible almost
compatibly with the $N=4$ unit.
The physical implication of kinks is discussed in the next section using the Wilson loop operator.
\par
Next, we examine the $U/t_{\rm eff}$ dependencies of the energy eigenvalues.
In Figs.~\ref{f5}(a) and \ref{f5}(b) those of the isolated $N=4$ unit are shown
for three choices of $\theta$.
The perturbation up to fourth order works well for $U/t_{\rm eff}\gtrsim 10$ for all cases,
and for smaller $U$ the perturbation energy falls off from the curve and starts to diverge.
\par
When combining these squares \ch{to} make a diamond chain or square lattice, 
the description of the spin model may seemingly become 
less accurate as we discard more higher order terms. 
However, the difference in the energy {\it per site}, $\Delta E/N$, 
between the spin model and the SOC Hubbard model, 
shown in the right insets of Fig.\ref{f5}(c,d), 
does not seem to depend that much on the lattice size; 
At $\theta=0$ the diamond chain or square lattice has smaller $\Delta E/N$, 
but it is opposite at $\theta=\pi$.  
\par
Importantly, the agreement in energies are good 
down to $U/t_{\rm eff}\gtrsim 5$ for all parameters or lattices we examined. 
In particular, at $U/t_{\rm eff}<10$ the role of fourth order energy becomes 
substantial as it shows opposite tendencies from the second order (lowered as $U/t_{\rm eff}$ becomes  smaller), and compensates for the upturn of second order energy.
The typical Mott transition takes place at around $U/t\sim 5$ and
most of the interesting material phases may lie in $U/t_{\rm eff}\gtrsim 5-10$.
Therefore, our results indicate that the obtained effective spin Hamiltonian is helpful in understanding
the underlying mechanism of the magnetism of the Mott insulating phase down to the vicinity of the Mott transition.
\par
Figures~\ref{f5}(c) and \ref{f5}(d) show separately the contribution
of different types of interactions to the perturbation energy.
For example, one finds that the two-body terms are dominant even at $U/t_{\rm eff}\sim 10$,
while the four-body \ch{terms} become increasingly important at $U/t_{\rm eff}\lesssim 10$.
The ring exchange term almost \ch{completely accounts for the four-body contribution at $\theta\sim 0$}, 
whereas, with increasing $\theta$, the other four-body terms give additional contribution 
and at $\theta=\pi$ they dominate.

\section{Summary and Discussion}
\label{sec:summary}
We derived the effective spin Hamiltonian using the strong coupling perturbative
expansion up to the fourth order, ${\cal H}^{(2)}+{\cal H}^{(4)}$,
from the half-filled Hubbard model with the spin-orbit coupling term.
The coupling constants of the interaction terms are compared between the second and fourth orders,
which showed that the fourth order terms develops at $U/t_{\rm eff}\lesssim 10$
and reaches almost half the second order ones at $U/t_{\rm eff}\lesssim 8$.
By diagonalizing the effective spin Hamiltonian and original Hamiltonian for four-site cluster and
for larger size ladders or square lattices, we find that the energies of the two Hamiltonian
agrees fairly well in this parameter range.
Particularly for the lattice clusters,
the energy eigenvalues of the two \ch{Hamiltonians
agree} very well even down to  $U/t_{\rm eff}\sim 5$.
\par
The effective spin Hamiltonian includes various types of terms at fourth order which are
mostly the combinations of Dzyaloshinskii-Moriya (DM),
Kaplan-Shekhtman-Aharony-Entin-Wohlman (KS), and ring exchange terms.
Although it is seemingly difficult to systematically understand the interplay between them,
the comparison of \ch{the expectation values} of these terms may give a clue to understanding the underlying mechanisms
of how the magnetic phases near the Mott transition may compete with each other \ch{with minimal set of terms that have dominant contribution to the full Hamiltonian}.
\par
Following Ref.[\onlinecite{Kawano2023}], let us expand the discussion on the quantum phases with finite SOC
using the SU(2) gauge that appeared as the hopping matrix in Eq.(\ref{eq:teff}),
defined as $U_{ij}=\mathrm{e}^{i(\theta/2)\bm{n}_{ij}\cdot\bm{\sigma}}$.
Here, for the spin quantization axis, we take the global spin coordinate common to all sites,
while the gauges depend on the choice of spin coordinate,
where a local gauge transformation can twist them independently.
However, the physical quantities do not depend on the choice of gauges,
and thus the gauge-invariant quantity can be an important clue to understand the effect of SOC.
The Wilson loop is a gauge-invariant quantity, which is the trace of the product of SU(2) gauges when hopping
around the plaquette, given as
\begin{align}
W_{\text{f}}&= \text{Tr}\big( U_{12}U_{23}U_{34}U_{41} \big)\equiv \text{Tr} \big( e^{i(\Phi_{\text{f}}/2) \bm m\cdot \bm \sigma}\big)
\notag\\
&=2\cos(\Phi_{\text{f}}/2).
\label{eq:wlf}
\end{align}
$\Phi_{\text{f}}\in [0, 2\pi]$ denotes the rotation angle and the three-dimensional unit vector $\bm m \in {\mathbb R}^3$
is the rotation axis, when the electron hops around the loop.
For type-C, $\bm n$ points in the $z$-direction so that the SU(2) gauge is reduced to the U(1) gauge
separately for the up and down spin electrons, and we find $W_{\text{f}}=2\cos(2\theta)$.
For type-N, $W_{\text{f}}=2(1-2\sin^4(\theta/2))$\cite{Sun2017}.
\par
Equivalently, for the spin Hamiltonian after perturbation,
the same argument applies.
By multiplying the rotation matrix of spin coordinates about axis-$\alpha_{ij}$,
given as $R^{\alpha_{ij}}(\theta_{ij})$ we obtain another Wilson operator,
\begin{align}
W_{\text{R}}&= \text{Tr}\big( R^{\alpha_{12}}(\theta_{12})  R^{\alpha_{23}}(\theta_{23})
  R^{\alpha_{34}}(\theta_{34})  R^{\alpha_{41}}(\theta_{41}) \big) \notag\\
&\equiv  1 + 2 \cos\Phi_{\text{R}}
\label{eq:wlr}
\end{align}
For type-C, the expression straightforwardly yields a $4\theta$ rotation about the $z$-axis
and gives $W_{\text{R}}=1+2\cos(4\theta)$.
For type-N, we find $W_{\text{R}}=4(1-2\sin^4(\theta/2))^2-1$.
These values are plotted in Figs.~\ref{f4}(e) and \ref{f4}(f) to compare with the energy eigenvalues as a function of $\theta$.
The location of kinks, $\theta=\pi/4, 3\pi/4$ for type-C and $\theta\sim 0.7$ for type-N,
coincide with the points where $W_{\text{f}}=-1$ and $W_{\text{R}}=0$ or equivalently, $\Phi_{\text{f}}=\Phi_{\text{R}}=\pi$.
These kinks emerge particularly within the insulating phase,
where the perturbative effective spin Hamiltonian provides an accurate description.
Moreover, the phase boundaries between the spiral and stripe phases
in the square lattice SOC Hubbard model (type-N) lie close to these points \cite{Kawano2023}.
Important features appear in the pyrochlore and kagome lattices:
for Wilson loops around the triangular unit, the condition $W_{\text{f}}=-1$ and $\Phi_{\text{f}}=\pi$ corresponds to
an emergent chiral symmetry in the band structure \cite{Nakai2023}, while
at $W_{\text{f}}=0$ and $\Phi_{\text{f}}=2\pi$ are associated with the emergence of a flat band \cite{Nakai2022}.
Although the direct physical implications of these quantities for the ground state remain unclear,
they provide insight into the gauge structures arising from SOC.
Indeed, the half-filled square lattice SOC Hubbard model at $\theta=\pi$ is equivalent to
the SU(2)-symmetric Hubbard model with the $\pi$-flux,
where $W_{\text{f}}=-2$ and $\Phi_{\text{f}}=2\pi$\cite{Kawano2023}.
Furthermore, the hole-doped phase diagram of this model \cite{Hodt2023}
exhibits various phases that may be understood through a combination of these effective models
and the role of hole degrees of freedom.
\par
Another important platform is the triangular lattice, where each plaquette consists of two triangles,
and fourth-order perturbation effects may play a crucial role.
In the half-filled Hubbard model on the $t-t'$ anisotropic triangular lattice,
which hosts a spin-liquid phase \cite{Morita2002,Yunoki2006,Szasz2021},
spin exchange interactions have been studied up to the twelfth order,
revealing that the ring-exchange term appears to be the dominant interaction in the spin-liquid phase\cite{Mila2010}.
Previously, the authors investigated the phase diagram of the Rashba SOC-Hubbard model on the triangular lattice and identified several types of small-skyrmion phases emerging around $\theta\sim \pi/2$ for $U/t_{\rm eff}=5-8$\cite{Makuta2024}.
In these regions, for type-N parameters, second-order perturbation terms
such as DM and KS interactions are dominant.
Additionally, fourth-order terms, including DM$\times$H, ring-KS, and the DM or KS-related terms can become comparably significant.
This suggests that the competition among these interactions gives rise to the observed small-skyrmion phases.
This explanation is consistent with [\onlinecite{Heinze2011}], which found the skyrmions with nanometer scales in Fe thin-layer on Ir(111) and contributed the origin of skyrmions to DM interaction and chiral four-spin interactions.
Such a small skyrmion in chiral magnets has also been reported recently\cite{khanh2024}.
\par
In systems that break inversion symmetry, many models suggest that skyrmions emerge due to anisotropy,
in addition to Heisenberg and Dzyaloshinskii-Moriya (DM) interactions.
It often happened that the easy-axis anisotropy $(S^z_i)^2$ terms or other anisotropic parameters
are added by hand and are parameterized freely in order to generate favorable skyrmion phases.
However, in our model derived from the Hubbard Hamiltonian, two distinct perspectives on skyrmion formation can be considered:
(1) Skyrmions may originate from the KSAEW terms in second-order Hamiltonian, which
{\it always exhibits the easy/hard-axis anisotropy in the same direction as the DM vector}.
(2) Skyrmions may arise from four-body interactions, which can take comparably large values as the two-body ones.
The distinction between these two mechanisms manifests in the phase diagram through variations in skyrmion size.
The KSAEW interactions favor smaller skyrmions; thus, as $U$ increases and second-order perturbation terms dominate,
the influence of KSAEW grows, leading to the formation of nano-skyrmions.
For instance, in the phase diagram at $U/t_{\rm eff}=8$,
a seven-site periodic skyrmion phase emerges due to this effect.
In contrast, four-body interactions do not inherently favor smaller skyrmions.
When they become dominant, larger skyrmions, such as the 27-site periodic flake skyrmions appear
in the phase diagram at $U/t_{\rm eff}=5$.
In the SOC Hubbard model, skyrmion size scales with $U$, reflecting the competition between these two different mechanisms. 
\par
Intuitively, to have a small skyrmion, or the 
shorter-pitch chiral state, 
one needs to increase DM interactions against the ferromagnetic interactions which compete and form such states. 
In the present framework, we do not have much room to afford ferromagnetic interactions, which should come out 
effectively in the systems off half-filling. 
However, we still find that the DM$\times$H fourth order term is indeed large 
in the region where we found such small skyrmions. 
\par
Since we have derived a complete set of spin exchange interactions up to fourth order, 
the present spin Hamiltonian can be directly applied to lattices of various geometries 
built from the given plaquette unit. 
Depending on the context, the effective models can then be simplified to 
include only the dominant terms, which may already capture the essential 
features of magnetism. 
As we demonstrated, certain interactions such as DM$\times$H and ring-KS or 
KS$\times$H$_2$ are substantially larger than others. 
and the future perspectives may be to clarify the role of these particular emergent terms.

\begin{acknowledgments}
	We thank Katsuhiro Tanaka and Masataka Kawano for discussions.
	R.M. was supported by a Grant-in-Aid for JSPS Research Fellow (Grant No. 23KJ0801).
	This work is supported by KAKENHI Grant No. JP21H05191 and 21K03440
	from JSPS of Japan.
\end{acknowledgments}

\appendix
%
\begin{figure}
  \includegraphics[width=0.45\textwidth]{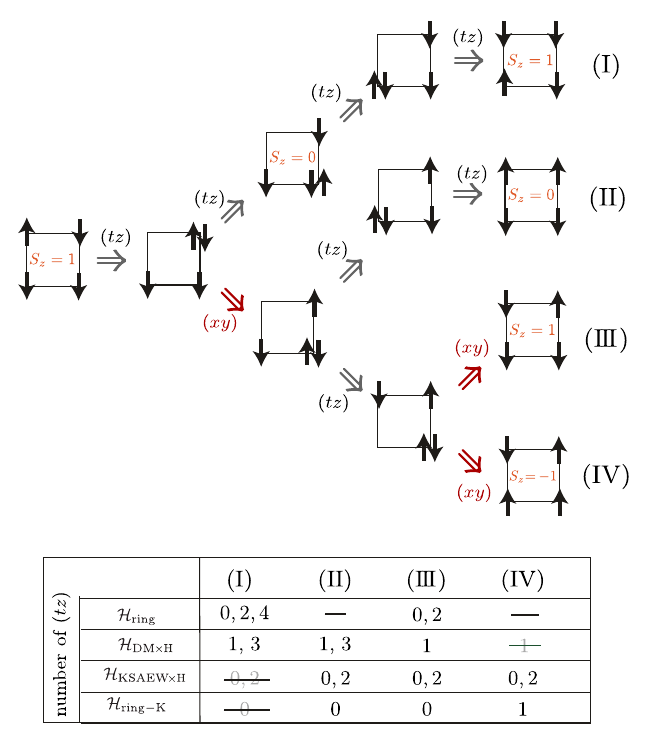}
  \caption{Fourth order perturbation processes (I)-(IV). $(tz)$ and $(xy)$ are those that preserves/changes
  the total $S_z$ when hopping. The lower panel shows the number of $(tz)$ process included in (I)-(IV) and the
  types of terms ${\cal H}_{\rm ring}$ , ${\cal H}_{\text{DM}\times\text{H}}$,
  ${\cal H}_{\text{KSAEW} \times \text{H}}$
  generated.}
  \label{f-Pros}
\end{figure}
\vspace{10mm}
\section{Origin of the exchange terms}
\label{app:exchange}
We discuss how the $\theta$-dependence emerges for each term listed in Tables I and II.
For this purpose, we have shown ${\cal J}_{\gamma}$ as functions of $c=\cos(\theta/2)$ and
$s=\sin(\theta/2)$ and kept the identity $c^2+s^2(=1)$ as a formula in the list.
Figure~\ref{f-Pros} show four different types of hopping processes (I)-(IV),
where we distinguish the constituent hoppings labeled as ($tz$) and ($xy$)
according to whether they come from $(t+i\lambda n_{z}\sigma_{z})$ which conserves total $S_{z}$,
or from $i\lambda(n_{x}\sigma_{x}+n_{y}\sigma_{y})$ which \ch{does not conserve} total $S_{z}$, respectively.
\par
Let us classify the $\theta$-dependent contributions
by parametrizing them as $\cos^k\theta/2\sin^{4-k}\theta/2$ with $k=0,1,2,3$.
Process (I) consists only of ($tz$) and does not change the spin orientation.
At each hopping we have two choices, $t$- and $\lambda$-terms,
which contains $\cos\theta/2$ and $\sin\theta/2$, respectively.
Therefore, assigning $k$ and $4\!-\!k$ hoppings to the former and latter, respectively,
all the contributions $k=0,1,2,3$ are allowed.
The ring exchange ${\hat h}_{\rm ring}^{\langle abcd\rangle}$ includes $k=0,2,4$
as it consists only of even numbers of $t$ and $z$-terms.
The ${\hat h}_{\text{KS}\times \text{H}}^{\langle abcd\rangle}$ has $k=0,2$
and ${\hat h}_{\text{ring-KS}}^{\langle abcd\rangle}$ has $k=0$.
However, the number of $k=0$ and 4 processes included in the process (I) is equivalent,
meaning that process I yield only
${\hat h}_{\rm ring}^{\langle abcd\rangle}$ and ${\hat h}_{\text{DM}\times \text{H}}^{\langle abcd\rangle}$.
\par
In process (II), the hopping changes the total $S^z$ from 1 to 0,
so that ${\hat h}_{\rm ring}^{\langle abcd\rangle}$ is excluded
but the other three terms have finite contributions.
Process (III) has two $(xy)$ terms so that setting number of $t$- and $z$-terms to be $k$ and $2-k$, respectively,
we find possible contributions from $k=0,1,2$.
The process conserves $S^z$ in total and all the four terms contribute, and among them,
only the ${\hat h}_{\rm DM \times \text{H}}^{\langle abcd\rangle}$ has odd $k$ contributions.
Finally, process (IV) that changes the total $S^z$ by 2 allows only
${\hat h}_{\text{DM}\times \text{H}}^{\langle abcd\rangle}$ and ${\hat h}_{\text{ring-KS}}^{\langle abcd\rangle}$.
\section{Roles of four-body terms}
\label{app:groundstate}
\begin{figure*}
  \includegraphics[width=0.8\textwidth]{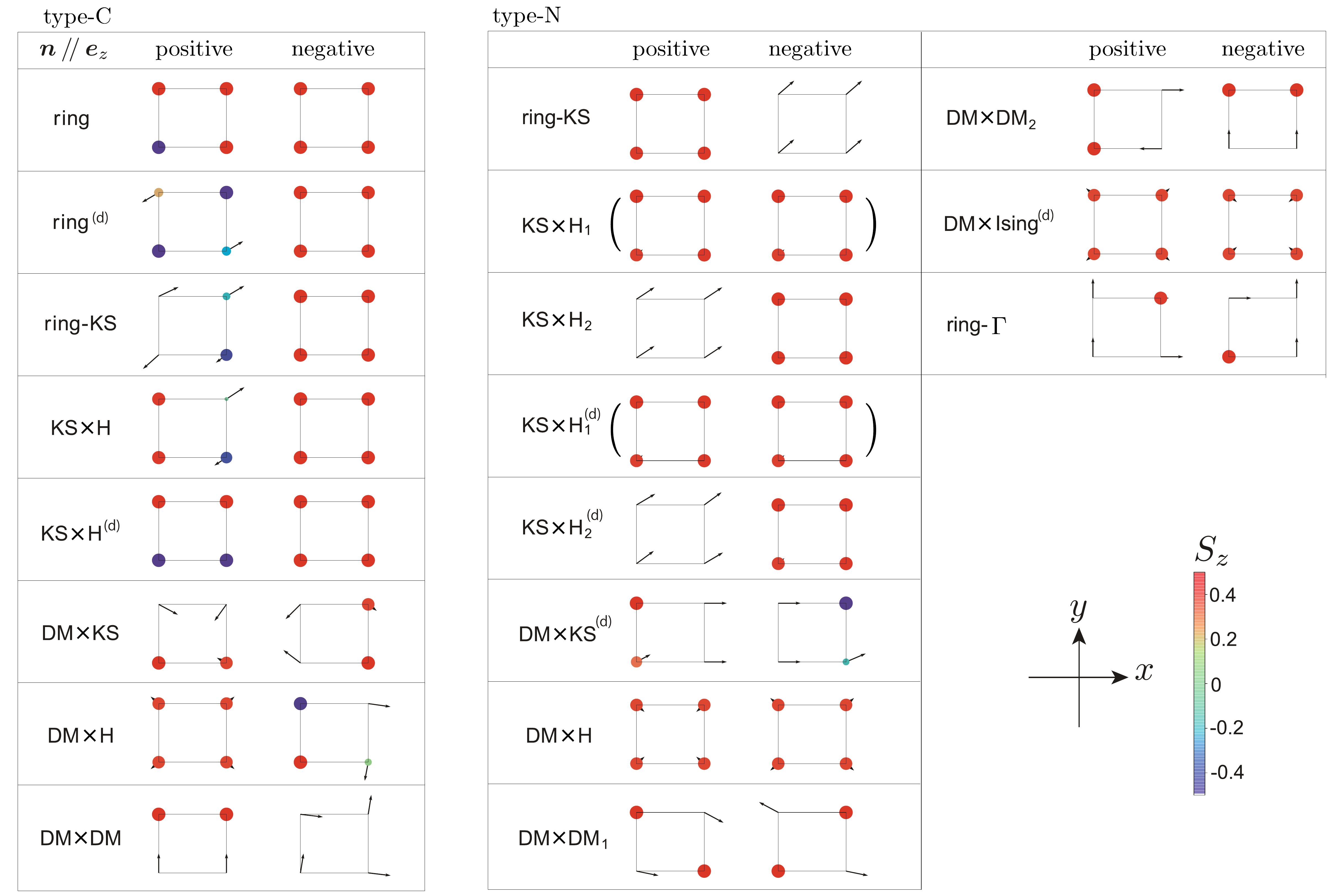}
  \caption{The lowest energy classical states
   for each of the term listed in Tables\ref{tab1} and \ref{tab2}.
  The spin configurations are depicted in the density plot of
  the amplitude of $\langle S^z\rangle$ and the vector
  for $\langle S^x\rangle, \langle S^y\rangle$ in the $xy$-plane.
 }
  \label{fGS}
\end{figure*}
We take a closer look at the role of each term.
Figure~\ref{fGS} shows the classical spin configuration on a unit plaquette
that is stabilized for the major terms that appeared in Tables I and II.
\par
In type-C, the two terms ring-KS and KS$\times$H, which are dominant at $\theta\sim \pi$,
favor both collinear antiferromagnet.
The three DM-related terms on the bottom take relatively large values
at $\pi/2 \lesssim \theta \lesssim \pi$ and may favor spiral or vortex-like structures.
The other terms basically favor ferromagnet.
\par
In type-N, the two dominant terms, ring-KS and KS$\times$H$_2$ favor in-plane
collinear structure, which may also cooperatively yield vortex or spirals when combined
with other interactions.
The four DM-related terms are relatively large at $\theta\lesssim \pi/2$, and
contribute to vortex or noncollinear types of in-plane structures.
\bibliography{ref}
\end{document}